\documentclass[12pt, preprint]{aastex}

\usepackage{graphicx}

\newcommand{\sbusr}{erg s$^{-1}$ cm$^{-2}$ sr$^{-1}$}

\newcommand{\ovi}{\ion{O}{6}}
\newcommand{\oiii}{[\ion{O}{3}]}
\newcommand{\oi}{[\ion{O}{1}]}

\newcommand{\kms}{km s$^{-1}$}

\newcommand{\fuse}{{\it FUSE}}

\newcommand{\iue}{{\it IUE}}

\newcommand{\rosat}{{\it ROSAT}}

\newcommand{\kpd}{KPD~0005+5106}

\slugcomment{Accepted for publication in the Astrophysical Journal}

\shorttitle{OVI around KPD0005+5106}
\shortauthors{Sankrit and Dixon}

\begin{document}

\title{A New Analysis of the \ion{O}{6} Emitting Nebula around
KPD~0005+5106\altaffilmark{1}}

\author{Ravi Sankrit\altaffilmark{2}}
\affil{SOFIA Science Center/USRA,
NASA Ames Research Center, 
Mail Stop 211-3, 
Moffett Field, CA 94035}
\email{rsankrit@sofia.usra.edu}

\and

\author{W. Van Dyke Dixon}
\affil{Department of Physics and Astronomy,
Johns Hopkins University, 
3400 N. Charles Street, 
Baltimore, MD 21218}
\email{wvd@pha.jhu.edu}

\altaffiltext{1}{Based on observations made with the NASA-CNES-CSA
{\it Far Ultraviolet Spectroscopic Explorer. FUSE}\/ is operated
for NASA by the Johns Hopkins University under NASA contract
NAS5-32985.}

\altaffiltext{2}{Visiting Research Scientist, Johns Hopkins University,
3400 N. Charles Street, Baltimore, MD 21218}

%%%%%%%%%%%%%%%%%%%%%%%%%%%%%%%%%%%%%%%%%%

\begin{abstract}

We present observations of \ovi~$\lambda$1032 emission around the
helium white dwarf \kpd\/ obtained with the \textit{Far Ultraviolet
Spectroscopic Explorer}.  Previously published data, reprocessed with
an updated version of the calibration pipeline, are included along with
new observations.  The recent upward revision of the white dwarf's
effective temperature to 200,000~K has motivated us to re-analyze all
the data.  We compare observations with photoionization models and find
that the density of the \ovi\ nebula is about 10~cm$^{-3}$, and that
the stellar flux must be attenuated by about 90\% by the time it
impinges on the inner face of the nebula.  We infer that this
attenuation is due to circumstellar material ejected by \kpd\ earlier
in its evolution.

\end{abstract}


%%%%%%%%%%%%%%%%%%%%%%%%%%%%%%%%%%%%%%%%%%

\section{Introduction}

The DO (helium-rich) white dwarf \kpd\ is the hottest known white
dwarf star.  Its effective temperature was measured to be
120,000~K by \citet{wer94}, and recently revised upward to 200,000~K
by \citet{wer07} based on the detection of \ion{Ne}{8} lines in its
spectrum.  Subsequently, \citet{wer08a} identified emission from
\ion{Ca}{10} in its UV spectrum, the highest ionization stage of
any element observed in a stellar photosphere.  Weak X-ray emission
at 1 keV detected with the \rosat\ PSPC has been interpreted as
evidence for a hot ($\sim 2 \times 10^6$ K), fast, perhaps shock-heated
wind, though it may be photospheric \citep{odw03, chu04, dra05,
wer08b}.  \citet{chu04} find that the star is surrounded by an
optical nebula, visible in \oiii~$\lambda$5007 emission,
roughly 3\arcdeg\ in diameter.  They find that the nebula contains
about 70~M$_{\sun}$ of material, and argue that this large mass
implies that the material must be interstellar rather than
circumstellar.  Closer in, \citet{ott04} discovered bright, diffuse
\ovi~$\lambda1032$ emission extending at least 3.5\arcmin\ from the star.

The \citet{ott04} study is based on data from 15 lines of sight
observed with the \textit{Far Ultraviolet Spectroscopic Explorer
(FUSE)} between 2001 and 2003.  \kpd\ was observed as a calibration
target throughout the \fuse\ mission, and spectra of half a dozen
additional sight lines are now available.  Meanwhile, the \fuse\
data-reduction software has evolved considerably, particularly with
the release of CalFUSE v3.  The revised effective temperature of
the white dwarf, the new data, and the improved calibration pipeline
motivated us to re-examine the nature of the \ovi\ emitting nebula
around \kpd\@.

%%%%%%%%%%%%%%%%%%%%%%%%%%%%%%%%%%%%%%%%%%

\section{Observations, Data Reduction, and Analysis}

\fuse\/ consists of four coaligned spectrographs. Two employ optics
coated with Al+LiF and record spectra over the wavelength range
990--1187 \AA; the other two use SiC coatings and are sensitive to
wavelengths below the Lyman limit. The four channels overlap between
990 and 1070 \AA.  Each channel has three apertures that
simultaneously sample different parts of the sky. The low resolution
(LWRS) aperture is 30\arcsec\ $\times$ 30\arcsec\ in size. The medium
resolution (MDRS) aperture is 4\arcsec\ $\times$ 20\arcsec\ and lies
about 3.5\arcmin\ from the LWRS aperture. The 1.25\arcsec\ $\times$
20\arcsec\ high-resolution (HIRS) aperture lies midway between the MDRS
and LWRS apertures. A fourth location, the reference point (RFPT), is
offset from the HIRS aperture by about 1\arcmin.  When a star is placed
at the reference point, all three apertures sample the background sky.
For a complete description of \fuse, see \citet{moo00} and
\citet{sah00}.

To identify observations for this study, we searched the Multimission
Archive at Space Telescope (MAST) for \fuse\/ observations of
\kpd\ obtained through the MDRS, HIRS, or RFPT apertures, for which the
LWRS aperture should sample the background sky.  Because its
sensitivity at 1032 \AA\ is more than twice that of any other channel,
we consider only data from the LiF 1A channel.  We use data
obtained in time-tag mode, which preserves arrival time and
pulse-height information for each photon event.  Our final sample,
consisting of data obtained along 21 lines of sight, is presented in
Table \ref{tab_observations}; the location of each sight line is
plotted in Figure~\ref{fig_map}.  Our sight lines now range over
more than 180\arcdeg\ around the star. 

The data were reduced using an implementation of CalFUSE v3.2, the
final version of the \fuse\/ data-reduction software package
\citep{dix07}, optimized for faint, diffuse emission.  Details are
given in \citet{dix06}, but we highlight a few points here:  the
program operates on one exposure at a time, producing a flux- and
wavelength-calibrated spectrum from each.  Background subtraction is
turned off.  Zero-point errors in the wavelength scale are corrected by
fitting a synthetic emission feature to the Ly$\beta$ airglow line.
The spectra from individual exposures are shifted to a heliocentric
wavelength scale and combined.  Finally, the spectra are binned by 14
pixels (53 \kms), a value chosen so that two binned pixels just
span a diffuse emission feature.

To improve the statistical significance of our results, we combine
spectra with the lowest signal-to-noise ratios with those of adjacent
sight lines, as noted in Table \ref{tab_intensities}.  The resulting
spectra are fit with a synthetic \ovi\ $\lambda 1032$ emission feature
using the non-linear curve-fitting program SPECFIT \citep{kri94}.  The
model consists of a top-hat shaped function 106 \kms\ in width,
representing the LWRS aperture, convolved with a Gaussian, representing
the instrument line-spread function ($\sim$ 25 \kms ) convolved with
the intrinsic emission-line profile.  Free parameters in the fit are
the level and slope of the background (assumed linear) and the
amplitude, central wavelength, and FWHM of the Gaussian.

Our model fits yield line intensities with $I/\sigma(I) \geq 2.9$
for 14 spectra.  The best-fit parameters are presented in Table
\ref{tab_intensities}.  (Note that the sight lines are ordered
anticlockwise, with those at radius 1.75\arcmin\ followed by those at
radius 3.5\arcmin.) Measured intensities range from 7.3 to 22.1
kLU.  (1\,LU = 1~photon~s$^{-1}$\,cm$^{-2}$\,sr$^{-1}$.)  A fit to
the sum of all 14 spectra yields an intensity of $15.5 \pm 0.8$
kLU, a line width of $30 \pm 10$~\kms, and a local standard of rest
velocity $v_{\rm LSR} = 11 \pm 2$ \kms.  Individual lines of sight
show considerable scatter about the mean velocity.  Several sight
lines have FWHM values $< 25$ \kms, indicating that their \ovi\
features are unresolved.  Two spectra show no evidence of \ovi\
$\lambda 1032$ emission; for them, we compute 3$\sigma$ upper limits
to the \ovi\ intensity by first modeling the linear continuum, then
increasing the strength of a model \ovi\ $\lambda 1032$ line until
$\chi^2$ rises by 9 (corresponding to a $3 \sigma$ deviation for
one interesting parameter; \citealt{avn76}).  Plots of selected
spectra and best-fit models are presented in Figure~\ref{fig_spectra}.
The asymmetric shapes of some model emission features are artifacts
of our sparse sampling of the model lines.

There are two caveats.  First, these observations are relatively
short.  To maximize the signal-to-noise ratio of our spectra, we use
all of the data, including that obtained during orbital day.  To test
for contamination by geocoronal emission \citep{she07}, we repeat the
analysis, using only data obtained during orbital night.  We find that,
in all cases for which the night-only spectrum has significant exposure
time ($> 1500$ s), the day-plus-night and night-only line intensities
agree within the night-only uncertainties.  Second, we do not attempt
to fit the fainter component of the \ovi\ doublet at 1038 \AA.  It is
only half as bright as the 1032 \AA\ line (in an optically-thin medium)
and may be blended with interstellar \ion{C}{2}* $\lambda 1037$ or
geocoronal \ion{O}{1} $\lambda 1039$.

The reddening toward \kpd, $E(B-V)$ = 0.06 (\S3.1), reduces the
observed \ovi\ intensity by $\sim$55\% \citep{fit99}.  The line
intensities presented in Figure~\ref{fig_map} and
Table~\ref{tab_intensities} are not corrected for this extinction.
The stellar spectrum shows strong absorption from molecular hydrogen
\citep{mur04}.  The H$_2$ absorption feature most likely to overlap
the \ovi\ $\lambda$1032 emission line is the Lyman (6,0) $R(4)$
transition at 1032.35~\AA\@.  Its velocity, $v_{\rm LSR} = -13$~\kms\
(J. Kruk, private communication), places it some 100~\kms\ from the
velocity of the summed \ovi\ emission features derived above.
Assuming that both the \ovi\ emitting and H$_2$ absorbing gas are
photoexcited and have low turbulent velocities, then their intrinsic
line widths should be about 10~\kms\ (a value consistent with the
30~\kms\ width of the summed \ovi\ feature, which includes the
instrumental line spread function).  The \ovi\ emission and H$_2$
absorption features are thus well separated.

Our results generally agree with those of \citet{ott04}.  Where
they differ, our intensities tend to be lower.  Examples are sight
lines S4044403 and M1070224 (map keys S43 and M24), which
\citeauthor{ott04} fit separately, obtaining intensities of $25 \pm
4$ and $25 \pm 7$ kLU, respectively.  Because the intensity of the
M24 feature is poorly constrained in our fits, we combine the two
spectra and obtain a best-fit intensity of $17.0 \pm 2.5$ kLU.  For
sight line M1070223, the difference is more extreme: our intensity
of $9.4 \pm 3.0$ kLU is half of the $18 \pm 8$ kLU reported by
\citeauthor{ott04}, though their value is poorly constrained.  Note
that, for most lines of sight, \citeauthor{ott04} use only data
from the nighttime portion of the orbit.

%%%%%%%%%%%%%%%%%%%%%%%%%%%%%%%%%%%%%%%%%%

\section{Modeling the Nebular Emission}

A star with an effective temperature of 200,000~K is expected to
produce a significant amount of high energy radiation, capable of
ionizing oxygen up to O$^{5+}$.  Thus, the \ovi\ emission we detect
around \kpd\ is most likely from gas photoionized by the star.  We
explore the conditions in the nebula by running photoionization
models.  The calculations, described in detail in \S3.2, were
performed with version 08.00 of Cloudy, last described by \citet{fer98}.

We require the models to reproduce two observational results: the
``typical'' \ovi\ intensity and the spatial extent of the \oiii\
emission detected by \citet{chu04}.  As we will show, these conditions
place strong constraints on the incident flux and on the density
of the emitting gas.  We do not seek to explain the distribution
of the \ovi\ emission, nor the variations in its intensity
(Figure~\ref{fig_map}), using these model calculations.  First, we
discuss the stellar parameters of \kpd, including its distance.
The latter is necessary for obtaining the spatial scales of the
emitting regions.

\subsection{Stellar Parameters}

In their initial analysis of \kpd, \citet{wer94} derive an effective
temperature $T_{\rm eff}$ = 120,000 K and a surface gravity $\log g =
7.0$.  Comparing these parameters with post-asymptotic giant branch
(post-AGB) evolutionary tracks, they estimate the star's mass to be
$M_* = 0.59 \; M_{\sun}$.  Comparing its V-band flux (V = 13.32;
\citealt{dow85}) with model predictions, they derive a distance to the
star of 270 pc.  At this distance, the 3.5\arcmin\/ radius of the outer
set of \fuse\/ sight lines (Figure~\ref{fig_map}) corresponds to a
physical extent of 0.27 pc.

The detection of \ion{Ne}{8} absorption features in the \fuse\/
spectrum of \kpd\ requires an upward revision of its effective
temperature.  \citet{wer07} find that the \ion{Ne}{8} lines are
best fit by models with $T_{\rm eff}$ = 180,000 -- 200,000 K and
$\log g = 6.5 - 7.0$.  Subsequently, \ion{Ca}{10} emission lines
were identified in the \fuse\/ spectrum, and fits to these lines
yield $T_{\rm eff}$ = 200,000 -- 220,000 K and $\log g = 6.2 - 6.5$
\citep{wer08a}.  For both species, the best-fit value of the surface
gravity rises with effective temperature.  \citet{wer07} point out
that the star's \ion{He}{2} lines, in both the UV and the optical,
are better fit by a model with $T_{\rm eff}$ = 200,000 K and $\log
g = 6.5$ than by the original $T_{\rm eff}$ = 120,000 K, $\log g =
7.0$ model.

K. Werner has kindly provided us with a synthetic spectrum of the
star.  It is based on a $T_{\rm eff}$ = 200,000 K and $\log g =
6.5$ model atmosphere composed of He, C, O and Ne, with C = 0.003,
O = 0.0006, and Ne = 0.01 by mass \citep{wer08a}.  J. Kruk has
provided the fully-reduced \fuse\/ spectrum of \kpd.  From these,
we can estimate both the star's interstellar reddening and its
distance.  We begin by scaling the synthetic spectrum according to
the reddening law of \citet{fit99} for various values of the color
excess $E(B-V)$.  We scale the result to match the star's V-band
flux.  Finally, we compare the predicted and observed flux at 1060
\AA\ in the \fuse\/ spectrum.  We find that the \fuse\/ data are
best fit by $E(B-V) = 0.06$, a value similar to the $E(B-V) = 0.07$
derived by \citet{dow85} from \iue\/ data.

As \citet{wer08b} point out, an effective temperature $T_{\rm eff}$
= 200,000 K and a surface gravity $\log g = 6.5$ place \kpd\/ on
the $0.7 \; M_{\sun}$ post-AGB evolutionary track of \citet{woo86}.
Adopting this mass and surface gravity, we derive a radius $R_* =
0.08 \; R_{\sun}$.  The scale factor required to make the reddened
model (at 5500 \AA) match the star's V-band flux is SF = $1.4 \times
10^{-23}$.  Given the star's radius and this scale factor, its
distance is simply $d^2 = \pi R_*^{\;2} / $SF \citep{kur79}, or $d
= 820$ pc.  At this distance, the 1.75\arcmin\/ radius of the inner
set of \fuse\/ sight lines corresponds to $\sim$ 0.42~pc ($1.3\times
10^{18}$~cm), the 3.5\arcmin\ radius of the outer set corresponds
to $\sim$ 0.84~pc ($2.6\times 10^{18}$~cm), and the 1.5\arcdeg\
radius of the \oiii\ nebula corresponds to $\sim$~21~pc ($6.6 \times
10^{19}$~cm).

We repeat the calculations using a second synthetic spectrum (also
provided by K. Werner) with $T_{\rm eff}$ = 180,000 K and $\log g
= 6.5$.  Scaling the model to match the observed V-band and 1060\AA\
fluxes, we derive a color excess $E(B-V) = 0.06$, which is the same
as for the hotter model.  Assuming the same stellar mass as before,
we derive a distance to the star of 780 pc.

Small changes in the surface gravity affect the stellar radius, but
do not significantly alter the overall shape of the emergent
continuum.  A star with $T_{\rm eff}$ = 200,000 K and $\log g =
6.2$ would have a radius of $0.11 \; R_{\sun}$, which is about 40\%
larger than that for $\log g = 6.5$.  The luminosity would therefore
be greater by about 90\% (since $L \propto {\rm R}^{2}$), and the
derived distance to the star would increase to 1.2 kpc.

\subsection{Cloudy Models}

Cloudy is a versatile code that can be used to calculate the
ionization structure and emission from photoionized gas under a
broad range of conditions.  Our models for the \ovi\ nebula detected
by \fuse\ are relatively simple.  We assume an ionization-bounded
slab of uniform density illuminated by the white dwarf.  The main
input parameters are the shape of the ionizing continuum, the
incident flux on the cloud surface, the density structure of the
gas, and the elemental abundances in the gas.  Calculations are
performed in one dimension assuming spherical symmetry; however,
the effective geometry can be made plane parallel by choosing a
large radius of curvature.

In our models, the abundances of elements are held fixed.  We use
the default solar values as specified in Cloudy: C and O from Allende
Prieto et al.\ (2001, 2002); N, Ne, Mg, Si and Fe from \citet{hol01};
and the rest from \citet{gre98}.  We take the $T_{\rm eff}$ = 200,000
K synthetic spectrum of \kpd\ as the shape of the incident continuum.
(In \S3.2.2 we use the $T_{\rm eff}$ = 180,000 K spectrum to study
how the results depend on the effective temperature of the star.)
The incident flux and the gas density are allowed to vary (\S3.2.3).

The physical distance from the star to the nebula scales with the
distance from the star to the earth.  The stellar distance is derived
by scaling the model spectrum to match the observed fluxes in the
V-band and at 1060~\AA\@.  Therefore the flux normalization at the
nebula does not depend directly on the distance between us and the
star, but rather on the stellar flux observed at earth and the
angular separation between the star and the inner face of the nebula,
which we take to be 1.7\arcmin.

\subsubsection{Initial model}

We first explore the case of the unattenuated stellar continuum
incident on the \ovi\ emitting nebula.  The inner face of the nebula
is assumed to be 1.7\arcmin\/ from the star, placing it just within
the inner set of \fuse\/ observations (clearly it can be no further
away).  This corresponds to a distance of $1.3\times 10^{18}$~cm
at 820 pc.  The normalization of the continuum may be described in
terms of the flux of photons with energy greater than or equal to
113.9~eV, which is the energy required to ionize O$^{4+}$ to O$^{5+}$.
(This ionization energy corresponds to a photon frequency of about
$2.76\times 10^{16}$~Hz.) From the synthetic spectrum of \kpd, we
find that the luminosity of such high-energy photons is $\sim
2.8\times 10^{46}$ photons~s$^{-1}$.  The flux of these photons at
the inner face of the nebula is $\sim 1.4\times
10^{9}$~photons~s$^{-1}$~cm$^{-2}$.  For this initial model, we
choose a hydrogen number density of 1~cm$^{-3}$, which is a typical
value for the interstellar medium (ISM).

The calculations predict that the \ovi\ emitting zone extends about
20~kpc, and that the \ovi\ $\lambda$1032 emissivity is about
$1.5\times 10^{-24}$~erg~s$^{-1}$~cm$^{-3}$ over most of the region.
\ovi\ $\lambda$1032 is a resonance line, and the Cloudy predictions
of emissivity take into account radiative transfer effects via the
escape probability formalism.  Although the escape probabilities
are calculated for a one-dimensional slab, we assume that the
predicted emissivity holds for our observed lines of sight, which
lie perpendicular to the direction of the stellar flux vector through
the model nebula.  Furthermore, we assume that the path length
through the nebula is about $4.2\times 10^{18}$~cm, which is roughly
equal to the extent of the \ovi\ emission in the plane of the sky
(about 5.7\arcmin, the distance between sight lines M08 and M23 in
Figure~\ref{fig_map}).  Using these values, we find that the expected
intensity from the model nebula is $\sim 5\times 10^{-7}$ \sbusr\@.
A color excess $E(B-V) = 0.06$ (\S\ 3.1) toward \kpd\ corresponds
to an extinction at 1032\AA\ of about a factor of 2 \citep{fit99}.
Therefore, the model predicts that the observed intensity would be
about $2.5\times 10^{-7}$ \sbusr, which equals about 13,000~LU\@.
This value is consistent with the \fuse\ observations.

Although this model reproduces the observed \ovi\ intensity, it does not
correctly reproduce the extent of the optical nebula.  In the model,
the \oiii\ emission arises beyond the \ovi, at a distance of between 20
and 32~kpc from the star.  This is about three orders of magnitude
larger than the $\sim 20$~pc radius of the \oiii\ nebula detected by
\citet{chu04}.

\subsubsection{$T_{\rm eff}$ = 180,000 K model}

The lower limit on the effective temperature of \kpd\ is 180,000
K\@.  We calculate a model using the $T_{\rm eff}$ = 180,000 K
synthetic spectrum to examine the sensitivity of the nebular
properties to the stellar effective temperature.  All other model
parameters remain unchanged from the initial model.

The luminosity of O$^{4+}$ ionizing photons is $\sim 1.4\times
10^{46}$ photons~s$^{-1}$, about half that of the 200,000 K stellar
spectrum.  Their flux is $\sim 6.8\times 10^{8}$~photons~s$^{-1}$~cm$^{-2}$
at the inner face of the nebula.

The \ovi\ zone extends about 12 kpc, just over half the distance
predicted by the 200,000 K model.  The total \ovi\ flux from the
180,000 K model nebula is only half that of the 200,000 K model.
However, the intensity in the inner few kpc is about 20\% higher.
This is because more of the O$^{5+}$ ions are ionized to O$^{6+}$
by the 200,000 K spectrum than the 180,000 K spectrum.  The \oiii\
emission arises at a distance of between 12 and 22 kpc from the
star.  This is still almost three orders of magnitude larger than
the observed radius of the optical nebula.

If the effective temperature of \kpd\ were at the lower limit allowed
by the \ion{Ne}{8} lines, the unattenuated flux impinging on gas
at typical interstellar densities would still produce an optical
nebula much larger than observed.  If the temperature were at the
higher limit (220,000 K), the discrepancy would be increased.  In
the rest of this paper, we consider only the best-fit 200,000 K
synthetic spectrum for \kpd.

\subsubsection{Fiducial model}

We ran a grid of models using the $T_{\rm eff}$ = 200,000 K synthetic
spectrum for the continuum shape, varying the incident flux
level and the density of the emitting gas.  The incident flux levels
ranged from the unattenuated stellar continuum used in the initial
model down to one-tenth this value.  (The latter corresponds to
$1.4\times 10^{8}$~O$^{4+}$ ionizing photons~s$^{-1}$~cm$^{-2}$.)
The hydrogen number densities ranged from 1~cm$^{-3}$ to 100~cm$^{-3}$.

For models in which the continuum is unattenuated, a gas density
$n_{\rm{H}}$ of about 40~cm$^{-3}$ is required to produce an \oiii\
nebula with a radius of about 20~pc.  In that case, however, the
predicted \ovi\ intensity is over 200 times the observed value.  As
the incident flux is lowered, the density required to reproduce the
observed radius of the \oiii\ nebula also decreases.  For each such
combination, the predicted \ovi\ intensity decreases with incident
flux and density, matching the observed values only when the incident
flux is $\sim$~10\% of the unattenuated value.  The corresponding
hydrogen density in the gas is 10~cm$^{-3}$.  We take this to be
the fiducial model for the nebula around \kpd\@.

In Figure~\ref{fig_model} we plot the predicted emissivities
of \ovi, \oiii, and H$\alpha$ against depth into the cloud.  The
average \ovi\ emissivity out to a few parsec is about $2\times
10^{-24}$~erg~s$^{-1}$~cm$^{-3}$.  This emissivity is slightly higher
than that predicted by the initial model and, under the same
assumptions used above, yields an observed intensity of 17,300~LU\@.
The \ovi\ nebula would be visible out to a few parsecs, i.e. beyond the
outer ring of \fuse\ observations.  The \oiii\ emissivity falls rapidly
beyond about 24~pc, and the H$\alpha$ emissivity extends slightly
beyond 26~pc.  The total \oiii\ to H$\alpha$ flux ratio is about 6.
This is consistent with the conclusion of \citet{chu04} that the
H$\alpha$ nebulosity associated with the nebula around \kpd\ is faint
compared with the \oiii\@.

The presence of photoionized \ovi\ emission around a hot white dwarf is
not unexpected.  As shown by our model calculations, a 200,000~K star can
maintain oxygen in this highly ionized state out to a distance of
several kiloparsecs for typical ISM gas densities and can produce the
observed \ovi\ $\lambda$1032 intensity.  (A similar point was made by
\citealt{pan84}, who showed that UV radiation from hot white dwarfs can
maintain the presence of C$^{3+}$ in the Galactic halo.) The puzzling
result is that, in order to produce the observed \ovi\ emission and an
\oiii\ emitting zone confined to a radius of $\sim20$~pc, the ionizing
continuum must be attenuated by about 90\% between its emergence from
the stellar surface and its incidence on the inner face of the nebula.
In the next section, we discuss how this attenuation may be explained
by the presence of circumstellar material around the white dwarf.

%%%%%%%%%%%%%%%%%%%%%%%%%%%%%%%%%%%%%%%%%%

\section{Discussion}

\subsection{Circumstellar material around \kpd}

The prior evolution of \kpd\ -- and in particular its mass-loss
history -- are unknown.  However, the star is expected to be
surrounded by material lost while it was on the asymptotic giant
branch.  We make the assumption that a shell of material surrounds
the star and examine the effect of such a shell on the stellar
continuum.  To do so, we calculate Cloudy models with the \kpd\
($T_{\rm eff}$ = 200,000 K) spectrum impinging on a shell and compare
the emergent continuum with the incident.

From the \citet{woo86} evolutionary track, we estimate that
KPD 0005+5106 left the asymptotic giant branch some 2000 years ago,
after ejecting its outer envelope.  We thus fix the inner radius of the
shell to be $1 \times 10^{17}$ cm, the distance
traveled in 2000 years by material moving at a typical PN expansion
velocity of 20 \kms. Assuming that envelope ejection took a few
thousand years, we fix the outer radius of the shell at three times the
inner radius. The density, assumed to be uniform, is treated as a
variable parameter. Abundances are fixed at the solar values. We find
that the continuum above 113.9 eV is attenuated by about 90\% when the
density of the gas in the shell is n$_{\rm H}$ = 3500 cm$^{-3}$. The
total mass in the shell (assuming that it covers 4$\pi$ steradians) is
then $\sim$ 0.4 $M_{\sun}$, comparable to the 0.2 $M_{\sun}$ typical of
planetary nebulae.

The shape of the emergent spectrum differs from that of the \kpd\
synthetic spectrum in several ways.  There is a sharp decrease at
the \ion{He}{2} edge at 54.4~eV from which the spectrum begins to
recover at $\sim 100$~eV\@.  While the O$^{4+}$ ionizing continuum
is reduced by 90\%, the continuum between 54.4~eV and 100~eV is
reduced by over 99\%.  The continuum between 13.6~eV and 54.4~eV
is reduced by only about 70\%.

We next calculate a Cloudy model of the nebula using this emergent
spectrum propagated out to the inner face of the nebula ($1.3\times
10^{18}$~cm from the star).  As described above, the shell density
is chosen such that the flux of O$^{4+}$ ionizing photons incident
on the nebula is the same as that for the fiducial model.  The
hydrogen density in the nebula is 10~cm$^{-3}$, also equal to that
in the fiducial model.  The difference in the shape of the continuum
has an impact on the predicted properties of the photoionized
nebula.

The resulting model has an \ovi\ zone that extends out to only about
2~pc, which is much smaller than that predicted by the fiducial
model (Figure \ref{fig_model}), but still extends beyond the outer
ring of \fuse\ observations (Figure \ref{fig_map}).  The \ovi\
intensity in the inner 1~pc (which encompasses the \fuse\ sight
lines), however, is about 3 times that of the fiducial model.  The
\oiii\ zone extends out to 45~pc, about twice that predicted by the
fiducial model (and thus about twice the observed extent of the
optical nebula).

The \ovi\ intensity and the extent of the \oiii\ zone can be brought
closer to the observed values if the flux emergent from the shell
is reduced by an additional 32\%, such that there are $\sim 9.5\times
10^{7}$~O$^{4+}$ ionizing photons~s$^{-1}$~cm$^{-2}$ impinging on
the nebula.  A model with this lower flux yields an \ovi\ zone
extending about 2~pc from the star (as above).  The \ovi\ intensity
in the inner 1~pc is within 5\% of that predicted by the fiducial
model, and the \oiii\ zone extends out to about 28~pc.

The gas density in the shell, its inner and outer radii, and the
abundances of elements are all unconstrained parameters of the
models.  Additionally, the nebular gas density may be adjusted.
Because there are no independent observational constraints on the
putative circumstellar shell around \kpd, more detailed calculations 
are not warranted.

We have shown (in \S3.2) that in order to explain the \ovi\ and
\oiii\ observations of the nebula around \kpd, the stellar flux
must be attenuated.  We have shown that this attenuation can be
achieved by a shell of gas with properties consistent with mass
loss from the star while on the asymptotic giant branch.  We have
further shown that the emergent spectrum is capable of photoionizing
a nebula and producing the observed \ovi\ and \oiii\ emission.
However, we do not make any attempt to model both the shell and the
nebular emission self-consistently.  A shell such as the one we
have assumed would produce its own emission.  To our knowledge,
there has not been a careful examination of the region around the
star at sufficiently high resolution to place strong limits on any
such emission.

\subsection{Spatial distribution of the \ovi\ emission}

It is clear from Figure~\ref{fig_map} that the \ovi\ emission has
an azimuthal structure, with the faintest emission from the southern
and northwestern edges of our sample and the brightest towards the
southwest.  The pattern is as clearly seen in Table~\ref{tab_intensities},
where the intensities exhibit a monotonic increase followed by a
monotonic decrease going around the inner and outer rings.  The
only exception is sight line M34.  The M34 value may be less reliable
than the others because it has the shortest exposure time and is
barely a $3\sigma$ detection.  If we ignore M34, then all the sight
lines outside of the cone between lines connecting the white dwarf
and M09 on one side and M24 on the other have \ovi\ intensities
less than or equal to 9.5~kLU\@.  Within this cone, all sight lines
have \ovi\ intensities greater than or equal to 14.5~kLU\@.  At the
edges of the cone, only the inner ring shows the brighter \ovi; M08
and M23 are fainter (Figure~\ref{fig_map}).

The observed distribution of the \ovi\ intensities could be due to
an azimuthal dependence of either (i) the gas density in the nebula
or (ii) the incident ionizing flux.  We examine the conditions
necessary for each of these alternatives, restricting ourselves to
the predictions from one-dimensional models with uniform gas density.

For a given incident flux, the ionization parameter (ratio of
ionizing photon density to gas density) decreases with increasing
density, while the emission measure increases with increasing
density.  Thus, there is some value of density at which the \ovi\
emissivity peaks.  Our fiducial model assumes a gas density of
10~cm$^{-3}$.  Raising the density to 15~cm$^{-3}$ reduces the \ovi\
emissivity by a factor of 1.5.  If the density is 20~cm$^{-3}$,
then the emissivity falls by a factor of 2.2.  These emissivities
yield observed intensities (under the path length assumptions used
in \S3.2) of 11.5~kLU and 7.9~kLU, respectively.  Higher-density
models also predict smaller and brighter \oiii\ nebulae than the
fiducial model does, but no obvious azimuthal structure is seen in
the \oiii\ images presented by \citet{chu04}.  A model with a density
of 1~cm$^{-3}$ predicts an \ovi\ emissivity that is half the value
of the fiducial model, but is ruled out because its \oiii\ zone
extends to 2~kpc\@.

If the gas density is fixed, then a lower incident flux will yield
fainter \ovi\ emission.  An incident flux 90\% that of the fiducial
model results in a factor of 1.4 drop in the \ovi\ intensity.  If
the incident flux is 80\% that of the fiducial model, then the
intensity drops by a factor of just over 2.  These values correspond
to observed intensities of 12.5~kLU and 8.5~kLU respectively.  The
\oiii\ intensity and the extent of the \oiii\ emitting
region are not strongly affected by changes in the incident flux
-- they change by factors of less than 20\% among these models.

Spatial variations in the flux incident upon the \ovi\ emitting
nebula could be achieved by moderate density variations in the
circumstellar shell around \kpd.  For example, in the shell considered
in \S4.1, if the density were 4000~cm$^{-3}$ rather than 3500~cm$^{-3}$,
then the O$^{4+}$ ionizing continuum would be attenuated by about
98\%, rather than 90\%.  The circumstellar shell is not an
opaque object casting a shadow, but a source of diffuse radiation,
with considerable angular extent when viewed from any point on the
inner face of the nebula.  Therefore we cannot simply infer the
azimuthal structure of this shell based on the observed distribution
of \ovi\ intensities.  Calculating the radiative transfer through
the circumstellar shell to derive the incident continuum at each
point in the nebula is beyond the scope of this work.

Of the two alternatives for explaining the observed azimuthal
distribution of \ovi\ emission, the latter, i.e., ionizing flux
variations, is the one that has less effect on the distribution of
\oiii\ emission.  Therefore, we suggest that it is the more
straightforward explanation.

One implicit assumption in all of our models is that the \ovi\ and
\oiii\ arise in a single nebula with uniform density.  This need
not be the case.  The mass loss of stars on the asymptotic giant branch
is a complex process; in the case of \kpd, up to 2~M$_{\sun}$ may
have been lost over a period of $10^4-10^5$~years prior to the final 
mass ejection that resulted in the circumstellar shell we have
inferred in \S4.1.  If such were the case, then it would be
necessary to consider (at least) an outer region of circumstellar
material and the interstellar medium as independent emitting regions.
One may then naively expect that all the \ovi\ arises in the former,
while the \oiii\ emission is dominated by the latter.  Without further
data, detailed explorations of such a scenario are premature.

%%%%%%%%%%%%%%%%%%%%%%%%%%%%%%%%%%%%%%%%%%

\section{Concluding Remarks}

In this paper, we have presented a new analysis of the \ovi~$\lambda$1032
emission around the hot helium-rich white dwarf \kpd\@.  At the
time of our earlier analysis \citep{ott04}, the star was believed
to have an effective temperature of 120,000~K and lie at a distance
of 270~pc.  Since then, the effective temperature
of the star has been revised \citep{wer07}, and consequently its
inferred distance has increased to 820~pc (\S3.1).

We have modeled our \ovi\ measurements and the \oiii~$\lambda$5007
emission around the star \citep{chu04} as a photoionized nebula.
We have used a synthetic spectrum of \kpd\ for the ionizing continuum
and treated the nebula itself as a uniform, one-dimensional slab.
Based on the comparison of models with data, we have inferred the
existence of a circumstellar shell between the star and the \ovi\
emitting nebula that attenuates the stellar flux by about 90\%.  We
find that the nebula has a density of $\sim 10$~cm$^{-3}$.

The distribution of \ovi\ intensities on the plane of the sky shows
a clear azimuthal structure.  This structure may be due to a variation
either of the gas density or of the incident flux.  Within the
framework of our simple models, we present the conditions under
which we may expect the observed \ovi\ intensity distribution for
each alternative.  We find that variations in the density would
affect the brightness and the extent of the \oiii\ nebula in ways
that probably would be observable, but which are not seen.
However, a more careful examination of the optical emission with
accurate photometry is required before any definitive statement can
be made.  The \ovi\ intensity is more sensitive to variations in
the incident flux, and the observed differences can be explained
without affecting the properties of the predicted \oiii\ emission.

Our results clearly highlight the need for a thorough examination
of the region around \kpd\@.  One possible tracer of high ionization
gas is [\ion{Ne}{5}]~$\lambda$3426.  Our Cloudy models predict that
the strength of this line is about 10 times that of \ovi~$\lambda$1032.
A photometric, narrowband imaging study of the nebula in \oiii\ and
[\ion{Ne}{5}] emission would yield diagnostic information about the
distribution of material and the ionization structure of the nebula.
We have inferred the existence of a circumstellar shell based on
the synthetic spectrum of \kpd\ and photoionization models for the
nebular emission.  Its existence is plausible considering the
evolutionary history of stars on the asymptotic giant branch.  A
direct observation of such a shell would be of great value.

%%%%%%%%%%%%%%%%%%%%%%%%%%%%%%%%%%%%%%%%%%

\acknowledgments

The authors thank Birgit Otte for initiating this
project and providing the software used to construct Figure~\ref{fig_map},
Klaus Werner for providing synthetic spectra used in their analysis
of \kpd, and Jeff Kruk for providing the star's fully-reduced \fuse\/
spectrum.  We acknowledge with gratitude the extraordinary efforts
of the \fuse\/ operations team to make this mission successful.
We thank the referee for comments that motivated an extension of
our analysis and a clearer presentation of our results.
This research has made use of the NASA Astrophysics Data System
(ADS) and the Multimission Archive at the Space Telescope Science
Institute (MAST).  STScI is operated by the Association of Universities
for Research in Astronomy, Inc., under NASA contract NAS5-26555.
Support for MAST for non-HST data is provided by the NASA Office
of Space Science via grant NAG5-7584 and by other grants and
contracts.  The curve-fitting program SPECFIT runs in the IRAF
environment.  The Image Reduction and Analysis Facility is distributed
by the National Optical Astronomy Observatories, which is supported
by the Association of Universities for Research in Astronomy (AURA),
Inc., under cooperative agreement with the National Science Foundation.
This work is supported by NASA grant NNX06AD30G to the Johns Hopkins
University, and by the Universities Space Research Association.

\textit{Facilities:} \facility{FUSE}.

%%%%%%%%%%%%%%%%%%%%%%%%%%%%%%%%%%%%%%%%%%%%%%%%%%%%%%%%%%%
%FIGURES

\begin{center}

\newpage
\begin{figure}[ht]
\includegraphics[width=6.0in]{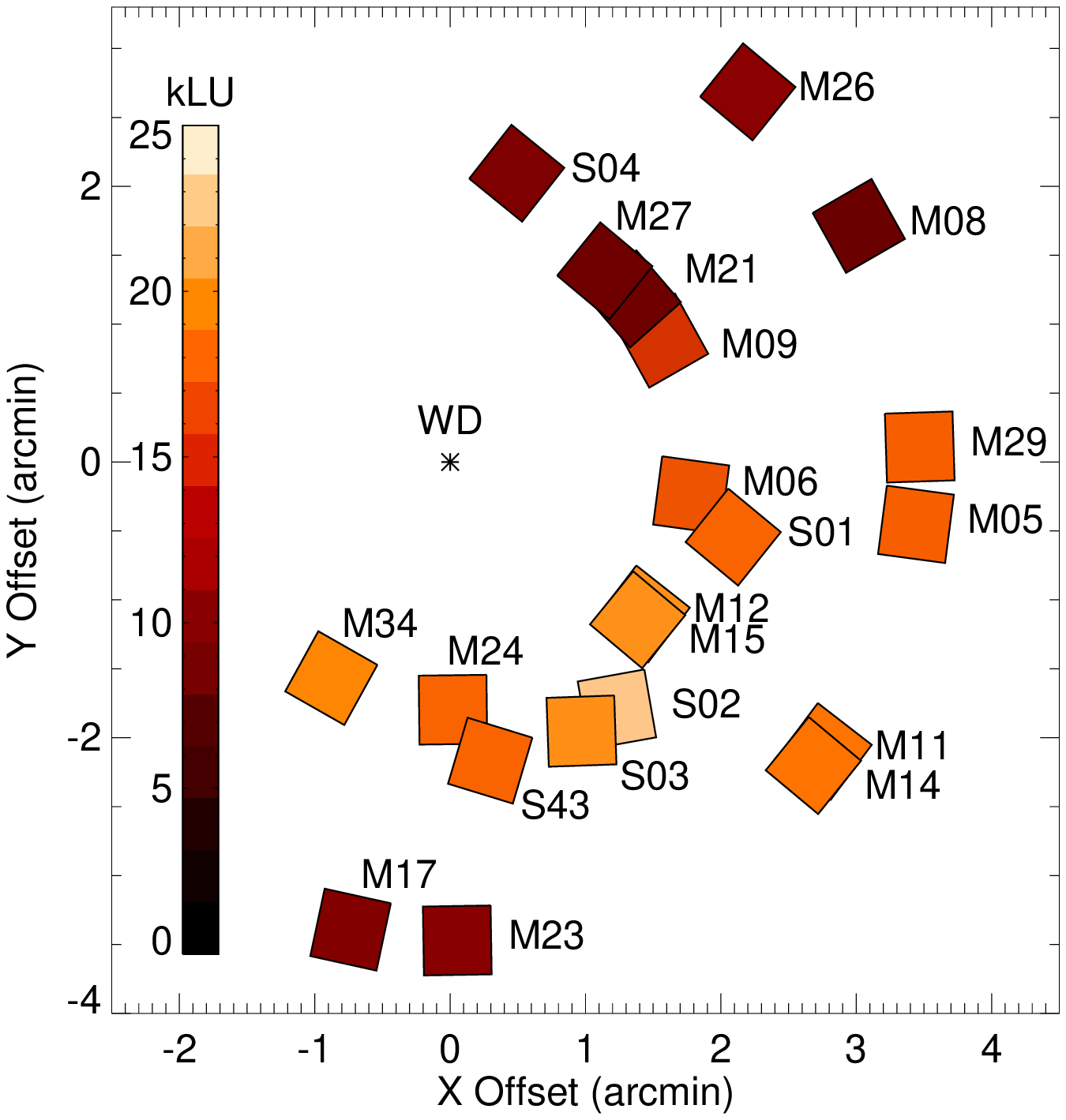}
\caption{Observations of diffuse \ovi\ emission around the hot
white dwarf KPD~0005+5106.  Colors represent the measured \ovi\/
$\lambda 1032$ intensity in units of kLU (1\,LU =
1~photon~s$^{-1}$\,cm$^{-2}$\,sr$^{-1}$) recorded through the \fuse\/
LWRS aperture.  For M17 and M26, the colors represent $3 \sigma$
upper limits.  The radius of the outer ring is 3.5 arcmin.  Labels
refer to the map keys defined in Table \protect{\ref{tab_observations}}.
North is up and east is to the left.
\label{fig_map} }
\end{figure}

\newpage
\begin{figure}[ht]
\includegraphics[width=6.0in]{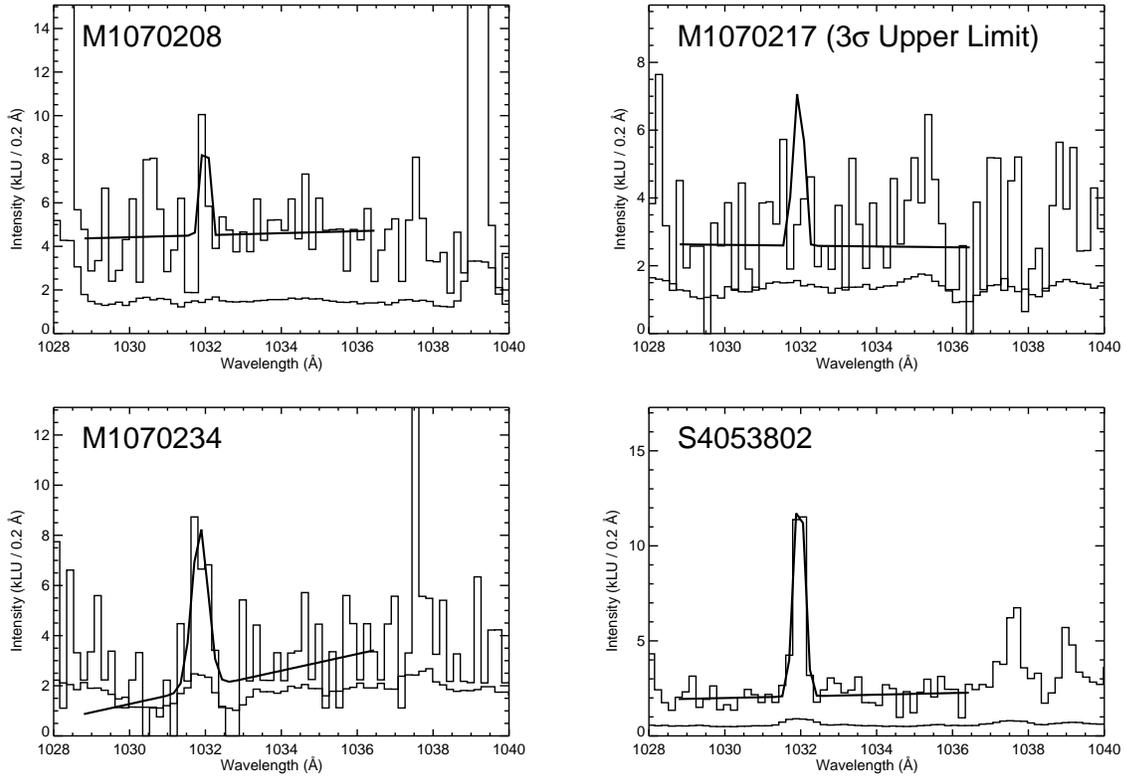}
\caption{Selected spectra showing the region around \ovi\
$\lambda 1032$.  The data are binned by 14 pixels, and best-fit
model and error spectra are overplotted.  For M1070217, the
model spectrum represents a $3 \sigma$ upper limit.  Interstellar
\ion{C}{2} * $\lambda 1037$ and \ovi\ $\lambda 1038$ are present in some
spectra, as are the geocoronal \oi \ $\lambda \lambda 1028, 1039$
lines.
\label{fig_spectra} }
\end{figure}

\newpage
\begin{figure}[ht]
\includegraphics[width=6.0in]{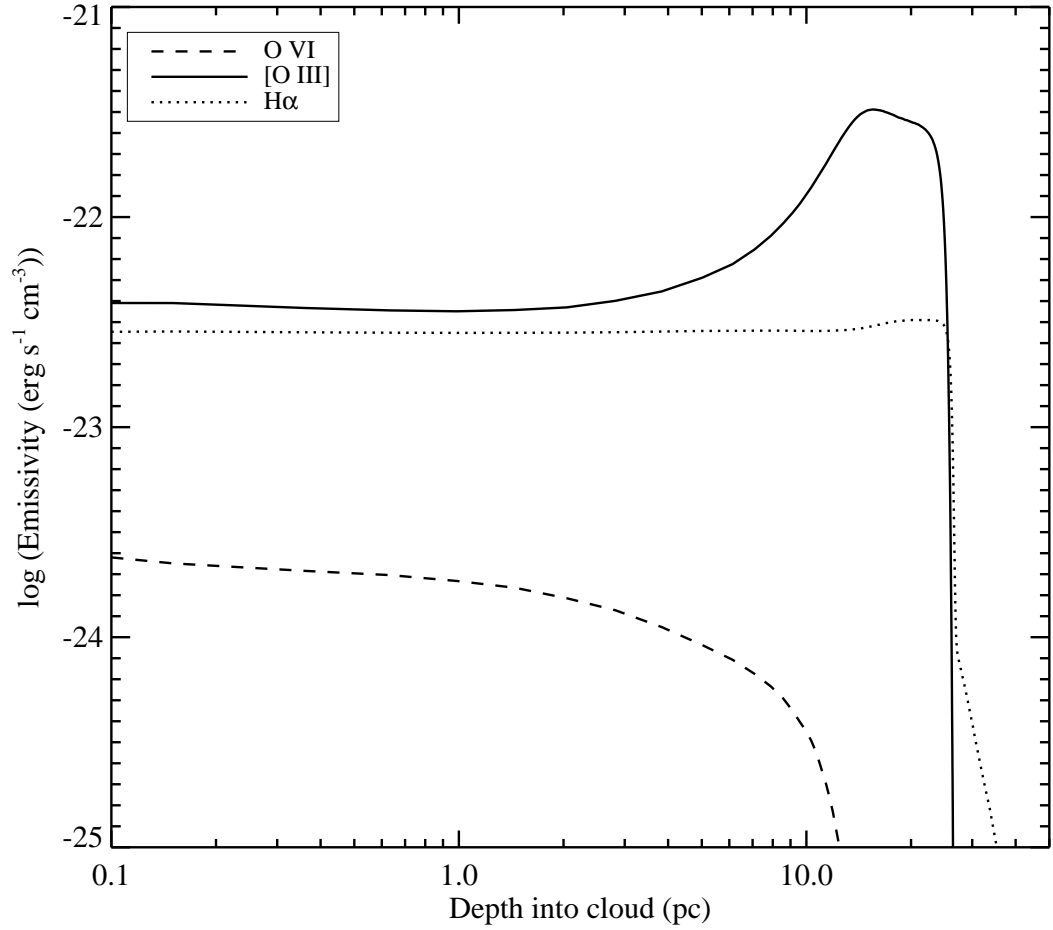}
\caption{Emissivities of selected lines predicted by the fiducial
model for the nebula.  The \ovi\ emissivity translates to an
observed intensity of 17,300~LU, consistent with the \fuse\ data.
The \oiii\ emission extends out to about 27~pc, consistent with
optical observations.
\label{fig_model} }
\end{figure}

\end{center}

%%%%%%%%%%%%%%%%%%%%%%%%%%%%%%%%%%%%%%%%%%%%%%%%%%%%%%%%%%%
%TABLES

\clearpage

\begin{deluxetable}{llccc}
\tablecolumns{5}
\tablewidth{0pt}
\tablecaption{Observations\label{tab_observations}}
\tablehead{
\colhead{Observation} & & \colhead{Position Angle\tablenotemark{a}} 
& \colhead{Distance} & \colhead{Map} \\
\colhead{ID} & \colhead{Date} & \colhead{(degrees)} 
& \colhead{(arcmin)} & \colhead{Key}}

\startdata
M1070205 & 2001 Sep 29 & 263 & 3.5 & M05 \\
M1070206 & 2001 Sep 29 & 262 & 1.8 & M06 \\
M1070208 & 2002 Sep 23 & 300 & 3.5 & M08 \\
M1070209 & 2002 Sep 23 & 299 & 1.8 & M09 \\
M1070211 & 2002 Oct 21 & 233 & 3.5 & M11 \\
M1070212 & 2002 Oct 22 & 232 & 1.8 & M12 \\
M1070214 & 2002 Oct 23 & 231 & 3.5 & M14 \\
M1070215 & 2002 Oct 23 & 230 & 1.8 & M15 \\
M1070217 & 2002 Dec 17 & 168 & 3.5 & M17 \\
M1070221 & 2003 Sep 07 & 311 & 1.8 & M21 \\
M1070223 & 2003 Dec 03 & 181 & 3.5 & M23 \\
M1070224 & 2003 Dec 03 & 181 & 1.8 & M24 \\
M1070226 & 2004 Jul 24 & 321 & 3.5 & M26 \\
M1070227 & 2004 Jul 24 & 321 & 1.8 & M27 \\
M1070229 & 2006 Sep 02 & 272 & 3.5 & M29 \\
M1070234 & 2006 Dec 26 & 151 & 1.8 & M34 \\
S4053801 & 2001 Oct 23 & 256 & 2.2 & S01 \\
S4053802 & 2004 Nov 23 & 215 & 2.2 & S02 \\
S4053803 & 2003 Dec 02 & 207 & 2.2 & S03 \\
S4053804 & 2004 Jul 23 & 346 & 2.2 & S04 \\
S4054403 & 2002 Dec 22 & 188 & 2.2 & S43
\enddata

\tablenotetext{a}{Position angle of LWRS aperture, measured east
of north, relative to the star.}
\end{deluxetable}

%%%%%%%%%%%%%%%%%%%%%%%%%

\begin{deluxetable}{lrcrrc}
\tablecolumns{6}
\tablewidth{0pt}
\tablecaption{Measured \ovi\ Intensities and Upper Limits\label{tab_intensities}}
\tablehead{
\colhead{Sight} & \colhead{Time\tablenotemark{a}} & \colhead{Intensity\tablenotemark{b}} 
& & \colhead{$v_{\rm LSR}$} & \colhead{FWHM\tablenotemark{d}} \\
\colhead{Line} & \colhead{(s)} & \colhead{(kLU)} & \colhead{S/N\tablenotemark{c}} 
& \colhead{(\kms)} & \colhead{(\kms)}
}

\startdata

M34     & 1836 & $18.8 \pm 6.2$    & 3.0 & $-13 \pm 23$ &  $120 \pm \phn60$ \\
S43/M24 & 6147 & $17.0 \pm 2.5$    & 6.9 & $4 \pm \phn7$ &  $\phn10 \pm \phn30$ \\
S03     & 9596 & $19.3 \pm 2.7$    & 7.3 & $ 8 \pm \phn7$ &  $\phn50 \pm \phn30$ \\
S02    & 14800 & $22.1 \pm 2.0$   & 11.1 & $18 \pm \phn5$ &  $\phn50 \pm \phn20$ \\
M12/M15 & 8029 & $19.4 \pm 2.8$    & 7.0  & $11 \pm \phn7$ & $ \phn50 \pm \phn40$ \\
S01    & 10703 & $17.0 \pm 1.8$    & 9.5 & $7 \pm \phn5$ &  $\phn\phn7 \pm \phn40$ \\
M06     & 3581 & $16.1 \pm 3.8$    & 4.2  & $13 \pm 11$ &  $\phn20 \pm \phn80$ \\
M09     & 4330 & $14.5 \pm 3.5$    & 4.1  & $8 \pm 16$ &  $120 \pm \phn50$ \\
M21/M27 & 6048 & $\phn7.6 \pm 2.3$ & 3.3 & $40 \pm 14$ & $ \phn10 \pm 110$ \\
S04     & 4111 & $\phn8.7 \pm 2.2$ & 4.0 & $4 \pm 11$ &  $\phn20 \pm 110$ \\
\\
M17  &   3160  & $ < 9.0 $\\
M23     & 2521 & $\phn9.4 \pm 3.0$ & 3.1  & $8 \pm 16$ & $ \phn10 \pm 140$ \\
M11/M14 & 6346 & $17.8 \pm 2.5$    & 7.2  & $8 \pm \phn7$ & $ \phn10 \pm \phn30$ \\
M05/M29 & 4457 & $16.7 \pm 2.8$    & 5.9 & $4 \pm \phn8$ &  $\phn10 \pm 100$ \\
M08     & 4214 & $\phn7.3 \pm 2.5$ & 2.9 & $26 \pm \phn9$ &  $\phn\phn3 \pm \phn30$ \\
M26  &   2806  & $ < 9.5 $

\enddata

\tablenotetext{a}{Total exposure time.}
\tablenotetext{b}{1\,LU = 1~photon~s$^{-1}$\,cm$^{-2}$\,sr$^{-1}$.}
\tablenotetext{c}{Statistical significance of quoted intensity, $I/\sigma(I)$.}
\tablenotetext{d}{Gaussian FWHM values include the smoothing imparted
by the instrument optics.  Values less than $\sim$ 25 \kms\ indicate
that the emission feature is unresolved.}
\end{deluxetable}

\end{document}